\documentclass[12pt]{article}
\usepackage{epsfig}
\usepackage{amsmath}
\usepackage{hhline}
\usepackage{amssymb}
\usepackage{times}
\usepackage{color}
\usepackage{cite}

\newlength{\dinwidth}
\newlength{\dinmargin}
\usepackage[ , ,bf,it]{caption}   
\usepackage[amssymb,thinspace,Gray]{SIunits}

\newcommand{\GeV} {\giga\electronvolt}
\newcommand{\MeV} {\mega\electronvolt}
\newcommand{\keV} {\kilo\electronvolt}
\newcommand{\eV}  {\electronvolt}
\newcommand{\mum} {\micro\meter}
\definecolor{dred}       {rgb}{0.00, 0.00, 0.00}  \newcommand{\dred}      {\color{dred}}

\begin{document}

\begin{titlepage}

  \noindent
  \begin{flushright}
    {\small
      DESY 09-028 \\
      SLAC-PUB-13551 \\
      ILC-NOTE-2009-049 \\
      July, 2009\\}
  \end{flushright}

  \vspace{1.0cm}

  \begin{center}
    \begin{Large}

      {\bf Polarimeters and Energy Spectrometers \\
        for the ILC Beam Delivery System} 

    \end{Large}
    \vspace{1.5cm}
    \begin{large}
      S.~Boogert$^1$, A.F.~Hartin$^2$, M.~Hildreth$^3$, D.~K\"afer$^2$, J.~List$^2$, 
      T.~Maruyama$^4$, K.~M\"onig$^2$, K.C.~Moffeit$^4$, G.~Moortgat-Pick$^5$, S.~Riemann$^2$, 
      H.J.~Schreiber$^2$, P.~Sch\"uler$^2$, E.~Torrence$^6$, M.~Woods$^4$ \\
    \end{large}

    \vspace{.3cm}
    $^1$Royal Holloway, University of London, UK \\
    $^2$DESY, Hamburg and Zeuthen, Germany \\
    $^3$University of Notre Dame, USA \\
    $^4$SLAC National Accelerator Laboratory, Stanford, USA \\
    $^5$IPPP, University of Durham, UK \\
    $^6$University of Oregon, USA \\
    
  \end{center}

  \vspace{1cm}

  \begin{abstract}
    Any future high energy $e^+e^-$ linear collider aims at precision measurements 
    of Standard Model quantities as well as of new, not yet discovered phenomena. 
    In order to pursue this physics programme, excellent detectors at the interaction 
    region have to be complemented by beam diagnostics of unprecedented precision. 
    This article gives an overview of current plans and issues for polarimeters and 
    energy spectrometers at the International Linear Collider, which have been designed 
    to fulfill the precision goals at a large range of beam energies from 45.6~\GeV{} at 
    the $Z^0$ pole up to 250~\GeV{} or, as an upgrade, up to 500~\GeV. 
  \end{abstract}

  \vspace{1.0cm}

  \begin{center}
    Submitted to {\em JINST} 
  \end{center}

\end{titlepage}

\newpage
\section{Introduction and Overview} 
\vspace*{-2mm}
The International Linear Collider (ILC) will open a new precision frontier, with beam 
polarization playing a key role in a physics program that demands precise polarization 
and beam energy measurements~\cite{role-pol}. The baseline configuration of the ILC, as 
described in the Reference Design Report (RDR)~\cite{rdr}, provides polarized electron 
and positron beams, with spin rotator systems to achieve longitudinal polarization at 
the collider interaction point (IP); upstream and downstream polarimeters and energy 
spectrometers for both beams; and the capability to rapidly flip the electron helicity 
at the injector, using the source laser. The possibility of fast positron helicity flipping 
is not included in the baseline configuration. A scheme for fast positron helicity flipping 
has been proposed~\cite{Moffeit1}.

The electrons will be highly polarized with $P(e^-) > 80\%$. Positrons will also be produced 
with an initial polarization $P(e^+) \sim 30-45\%$, which can be upgraded to 60\%. Even the 
small positron polarization expected in the inital setup can be used with great benefit for 
physics measurements if the possibility of fast helicity flipping of the positron spin is 
also provided. Excellent polarimetry for both beams, accurate to $\Delta P/P = 0.25\%$, is 
planned~\cite{role-pol,Aurand}.  Polarimetry will be complemented by $e^+e^-$ collision data, 
where processes like $W^{\pm}$ pair production can provide an absolute scale calibration for 
the luminosity-weighted polarization at the IP, which can differ from the polarimeter 
measurements due to depolarization in collision.

Precise beam energy measurements are necessary at the ILC in order to measure particle 
masses produced in high-rate processes. Measuring the top mass in a threshold scan to 
order 100~\MeV{} or measuring a Standard Model Higgs mass in direct reconstruction to 
order 50~\MeV{} requires knowledge of the luminosity-weighted mean collision energy 
$\sqrt s$ to a level of\linebreak $(1-2)\cdot10^{-4}\;$~\cite{role-pol, Aurand}. 
Precise measurements of the incoming beam energy are a critical com\-po\-nent to measuring 
the quantity $\sqrt s$ as it sets the overall energy scale of the collision process.

The baseline ILC described in the RDR provides collider physics with beam energies in the 
range 100-250~\GeV.  Precise polarization and energy measurements are required for this 
full energy range.  The ILC baseline also provides for detector calibration at the $Z$-pole 
with 45.6~\GeV{} beam energies.  However, the RDR does not require accurate polarimetry or 
energy spectrometer measurements at the $Z$-pole.  A proposal to modify the baseline ILC to 
require precise polarimetry and energy measurements at $Z$-pole energies was made at the 
{\it Workshop on Polarization and Beam Energy Measurements at the ILC}, held in Zeuthen in 
2008~\cite{Aurand}. The motivation for this includes polarimeter and energy spectrometer 
calibration, and physics measurements to improve on $Z$-pole results from LEP and SLC. 
The downstream polarimeter described in the RDR is expected to perform well at the 
$Z$-pole, while for the upstream polarimeter the necessary changes 
{\dred are described in this paper and} 
will be included in the next update of the ILC baseline design. 
For energy measurements, the downstream energy spectrometer should perform well 
while the upstream spectrometer needs further evaluation {\dred as to} how accurately the 
lower chicane magnetic fields can be measured.

\begin{figure}[h!]
  \hspace*{-2mm}\epsfig{file=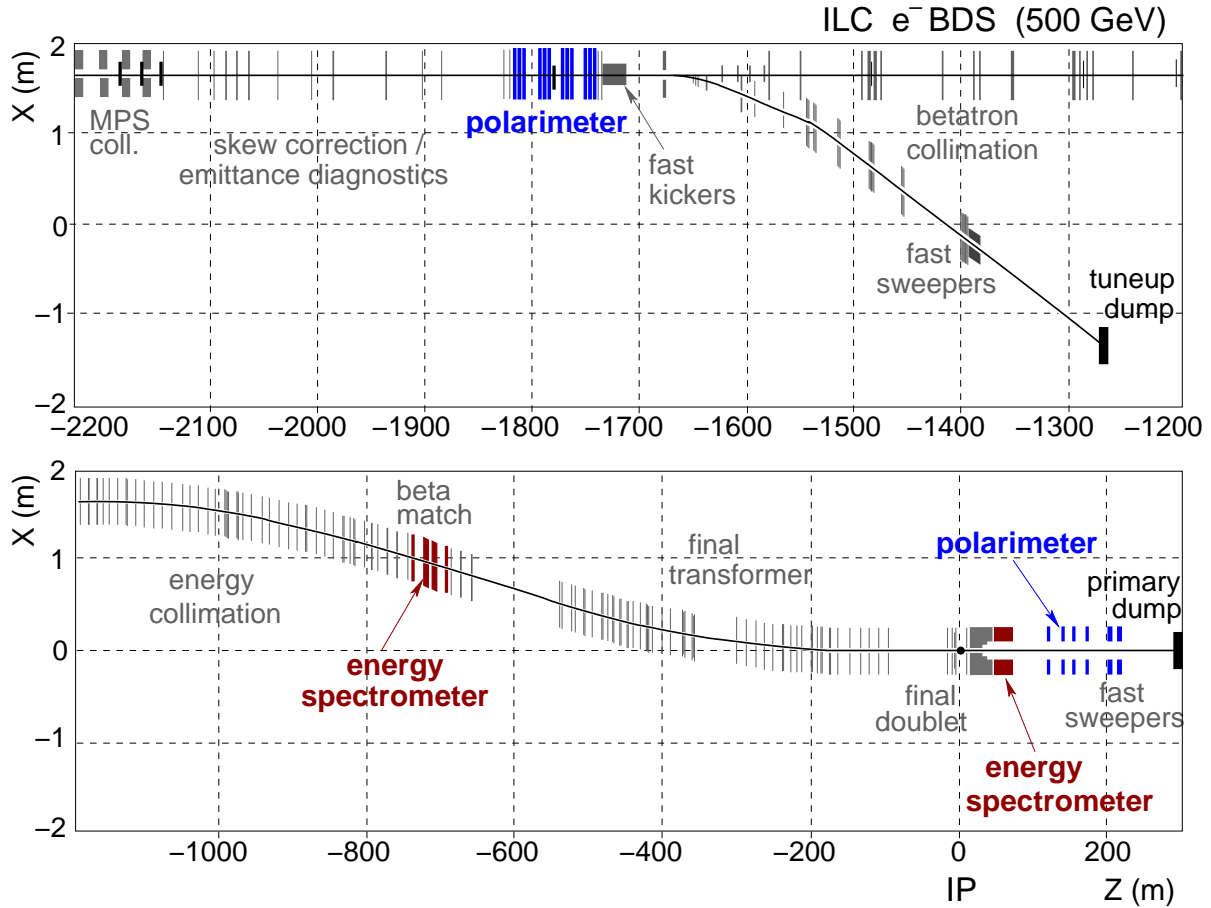, width=1.02\linewidth}
  \vspace*{-7.5mm}
  \caption[Beam Delivery System]{\it Beam Delivery System {\dred(BDS)} as described in the RDR. 
    The upper part shows the region from 2200~m to 1200~m upstream of the $e^+e^-$~IP, 
    including the polarimeter chicane at 1800~m. The lower part shows the region from 
    1200~m upstream to 400~m downstream of the IP, including the upstream energy 
    spectrometer at 700~m as well as the extraction line energy spectrometer and 
    polarimeter around 100~m downstream of the IP located at $z=0\;$m.}
  \vspace*{-1mm}
  \label{fig:BDS} 
\end{figure}

The locations of the polarimeters and energy spectrometers in the Beam Delivery System (BDS) 
as forseen in the RDR are shown in Figure~\ref{fig:BDS}. 
Data from the polarimeters and spectrometers must be delivered to the detector data 
acquisition system in real time to be logged and to permit fast online analysis. 
On the other hand, fast online analysis results must also be provided to the ILC 
control system for beam tuning and diagnostics. 

\enlargethispage{2mm}
{\dred 
  The purpose of this paper is to describe the polarimeters and energy spectrometers 
  which have been designed in order to fulfil the precision requirements over the 
  entire range of beam energies from 45.6~\GeV{} up to 500~\GeV.
  Section~\ref{sec:polarimetry} provides an overview of the functional prin\-ciple 
  of the BDS polarimeters, discusses possible choices of Cherenkov detectors and the 
  layouts of both magnetic chicanes, as well as the challenges caused by the forseen 
  14~mrad crossing angle of electron and positron beams at the IP.
  Section~\ref{sec:energy} presents an overview of the layout and technology foreseen 
  for the upstream and downstream energy spectrometers, while also provi\-ding a short 
  subsection on alternative methods for beam energy measurements.
}

\section{Polarimetry}\label{sec:polarimetry}
\begin{figure}[h!]
  \setlength{\unitlength}{1.0cm}
  \begin{picture}(12.0, 6.0)
    \put( 0.00,-0.10)  {\epsfig{file=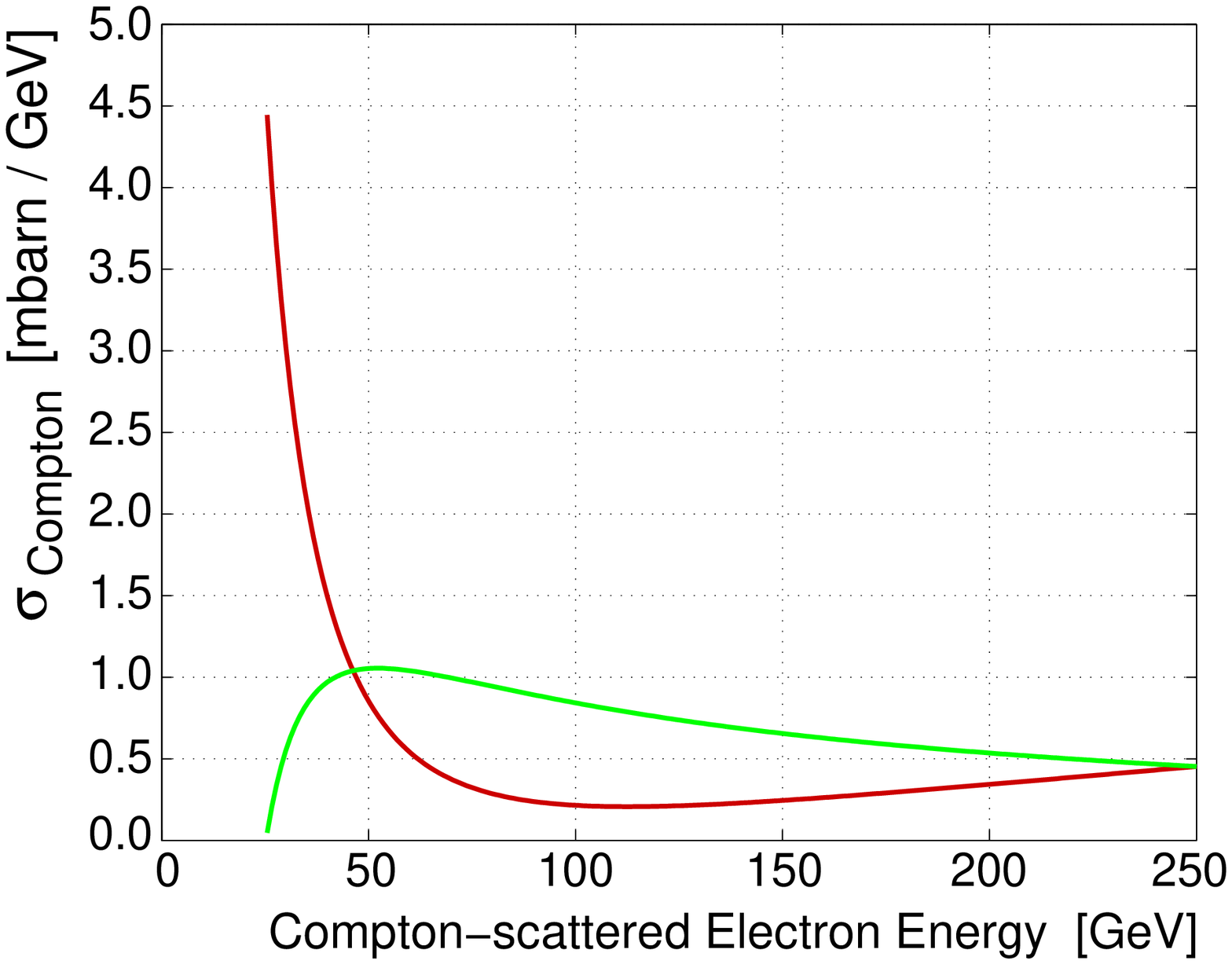,    bb=22 175 553 590, clip= , width=0.50\linewidth}}
    \put( 8.00,-0.10)  {\epsfig{file=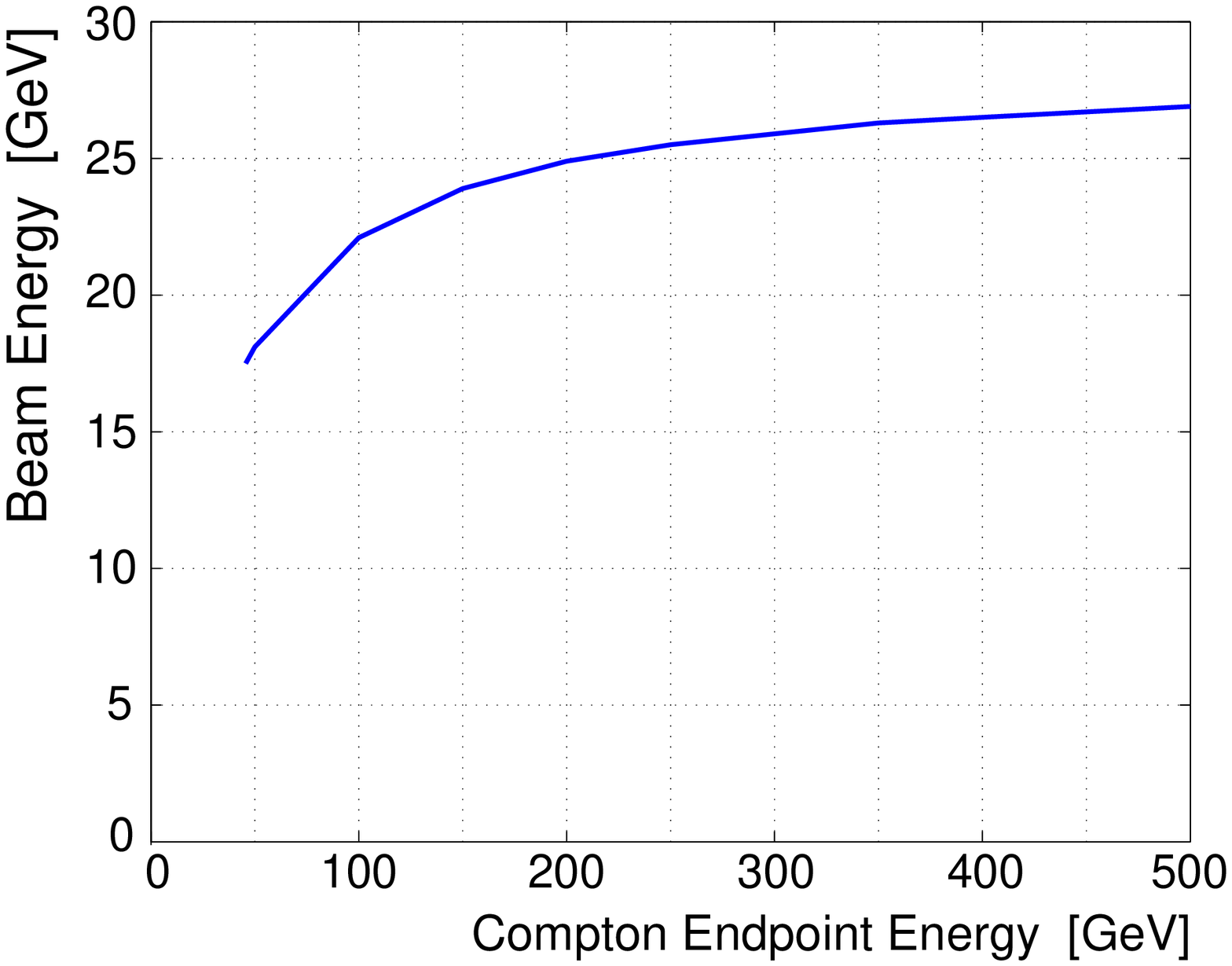, bb=22 175 553 590, clip= , width=0.50\linewidth}}
    \put( 2.00, 5.50)  {(a)}
    \put( 9.80, 5.50)  {(b)}
  \end{picture}
  \caption[Compton Cross Section] {\it 
    {\dred(a)} Compton differential cross section versus scattered electron energy for 
    same ({\dred black/}red curve) and opposite ({\dred grey/}green curve) helicity configuration 
    of laser photon and beam electron. The beam energy is 250~\GeV{} and the laser 
    photon energy is 2.3~\eV. 
    {\dred (b)} Compton edge energy dependence on the beam energy.}
  \label{fig:Compcross} 
\end{figure}

\vspace*{-2mm}
Both upstream and downstream BDS polarimeters will use Compton scattering of high power lasers 
with the electron and positron beams~\cite{role-pol, rdr}. Figure~\ref{fig:Compcross} shows the 
Compton cross section versus scattered electron energy for 250~\GeV{} beam energy and 2.3~\eV{} 
photon energy. There is a large polarization asymmetry for back-scattered electrons near 25.2~\GeV, 
the Compton edge energy.  The large asymmetry and the large difference between the Compton edge 
and the beam energy facilitate precise polarimeter measurements.  The Compton edge does not change 
significantly for higher beam energies; this dependence is also shown in Figure~\ref{fig:Compcross}.  

A spectrometer with segmented Cherenkov detectors that sample the flux of scattered 
electrons near the Compton edge will be used to provide good polarization measurements 
with high analyzing power.  Compton polarimetry, utilizing measurements of back-scattered 
electrons near the Compton edge, is chosen as the primary polarimetry technique 
for several reasons: \\[-6mm]
\begin{itemize}
  \item The physics of the scattering process is well understood {\dred from} QED, 
    with radiative corrections of less than 0.1\%~\cite{Swartz}; 
  \item Detector backgrounds are easy to measure and correct for using ``laser off'' pulses; 
  \item Compton-scattered electrons can be identified, measured and 
    isolated from backgrounds using a magnetic spectrometer; 
  \item Polarimetry data can be taken parasitic to physics data; 
  \item The Compton scattering rate is high and small statistical errors can be achieved 
    in a short amount of time (sub-1$\%$ precision in one minute is feasible); 
  \item The laser helicity can be selected on a pulse-by-pulse basis; and
  \item The laser polarization is readily determined with $0.1\%$ accuracy.
\end{itemize}
{\dred 
  It is expected that a systematic precision of $\Delta P/P = 0.25\%$ or better can be 
  achieved with the largest uncertainties coming from the analyzing power calibration 
  and the detector linearity, both of which are in the range of 0.1\% to 0.2\%~\cite{List}.
}

\begin{figure}[!h]
  \hspace*{-0.2mm}{\epsfig{file=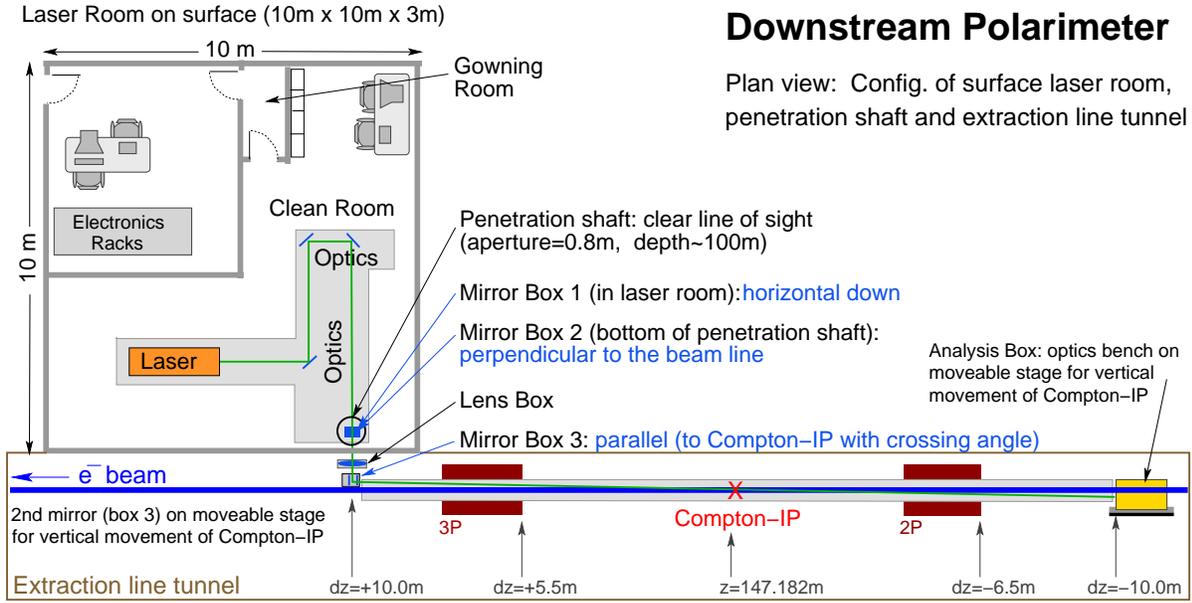, width=1.00\linewidth}} 
  \caption[Polarimeter Laser Room] {\it 
    Proposed configuration of laser room, penetration shaft 
    and extraction line layout for the downstream Compton polarimeter.}
  \label{fig:LaserRoom}
\end{figure}

Each polarimeter requires a laser room on the surface with a transport line to the 
beamline underground. A configuration proposed for the extraction line polarimeter 
is shown in Figure~\ref{fig:LaserRoom}. A similar configuration is planned for the 
upstream polarimeter. 
{\dred 
  The entire layout of the laser room and penetration shaft is conceptual and still 
  needs to be optimised to keep the radiation in the laser room below required levels 
  even for a worse case scenario of beam loss. 
  It is possible, for example, to pack the 80~cm wide penetration shaft with neutron 
  absorbing material at the top and bottom, which (togehter with its depth of about 100~m) 
  would reduce the neutron levels to a safe level in the surface laser room. 
}

\begin{figure}[h!]
  \center{\epsfig{file=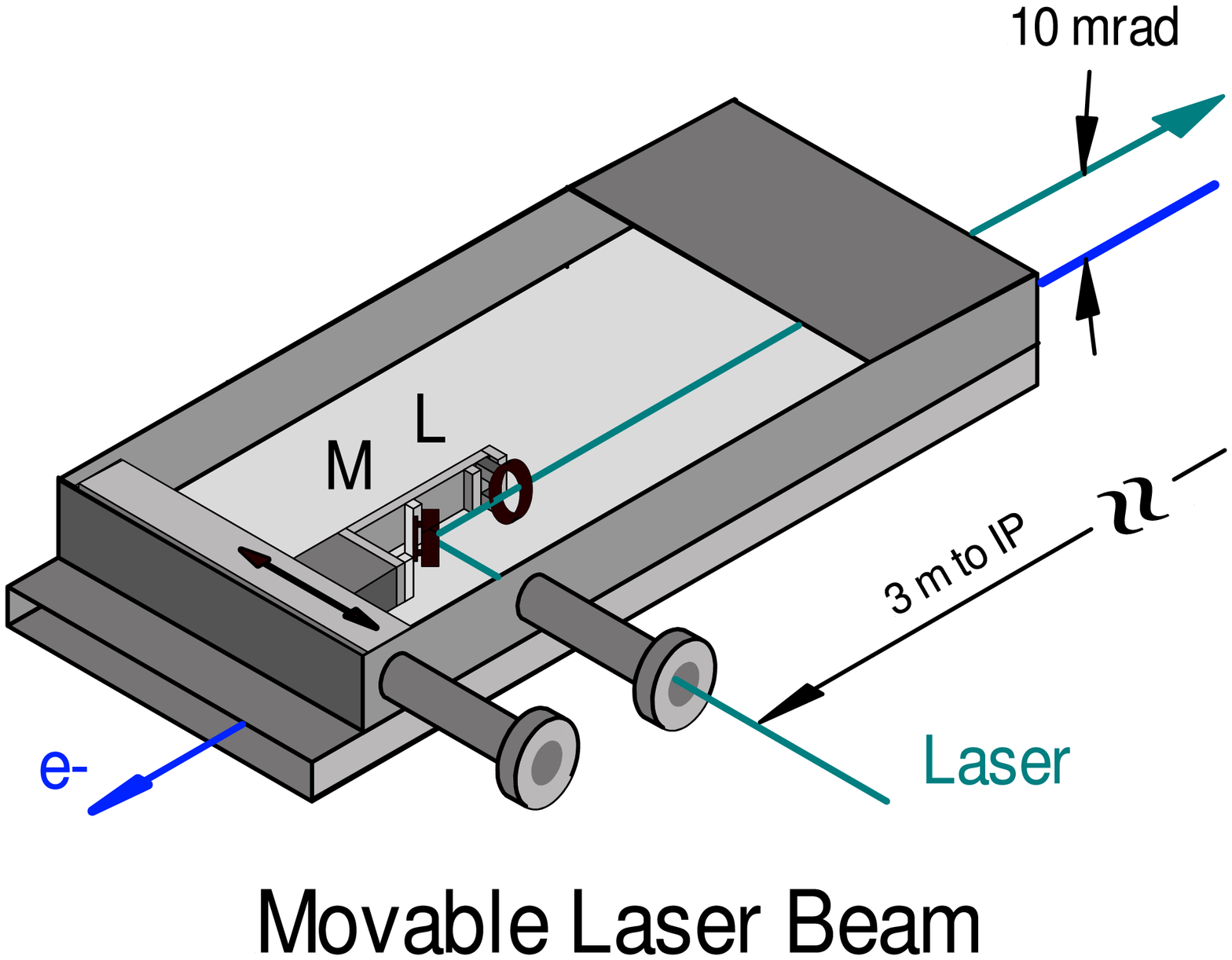, width=0.34\linewidth, angle=-90, clip=}}
  \caption[Movable Laser Optics] {\it 
    Movable mirror and lens focussing the laser onto the electron beam.}
  \label{fig:pol_upst_mirror} 
\end{figure}

The polarimeters employ magnetic chicanes with the parameters shown in Table~\ref{tab:polchicane}.
{\dred 
  The parameters where chosen such that a safe distance from the main beam can 
  be ensured for the Cherenkov detectors while simultaneously avoiding emittance 
  blow-up and keeping the cost at minimum.
  Both chicanes have been designed to spread the Compton spectrum horizontally 
  over about 20~cm, while allowing to keep a relatively low dispersion of 20~mm 
  at the mid-chicane point even for a beam energy of $E_b=250$~\GeV. 
  The spread of the Compton spectrum does not change and is independent of the beam 
  energy if the magnetic field is kept constant. This leads to a stable position 
  distribution of fixed shape and location at the surface of the Cherenkov detector. 
}
However, the Compton IP moves laterally with the beam energy. 
Figure~\ref{fig:pol_upst_mirror} shows a setup to adjust the laser accordingly, 
resulting in a maximal dispersion of about 12~cm at $E_b=45.6$~\GeV{} and a minimal 
dispersion of about 1~cm at $E_b=500$~\GeV. 
\begin{table}[h!]
  \renewcommand{\arraystretch}{1.15}
  \begin{tabular*}{\textwidth }{@{\extracolsep{\fill}}l crl crl}
    \hline\hline 
    Chicane Parameters          & \multicolumn{3}{c}{Upstream\quad Polarimeter} & \multicolumn{3}{c}{Downstream\quad Polarimeter} \\
    \hline\hline 
    Chicane Length [m]          & & \multicolumn{2}{c}{ 74.6 }                    & & \multicolumn{2}{c}{ 72.0 } \\
    Number of magnets           & & \multicolumn{2}{c}{ 12   }                    & & \multicolumn{2}{c}{  6   } \\
    Magnet Length [m]           & & \multicolumn{2}{c}{  2.4 }                    & & \multicolumn{2}{c}{  2.0 } \\
    \hline 
    \quad                       & &   \quad  &   \quad                            & &    0.4170   &  (1P, 2P)    \\
    Magnetic Field [T]          & &  0.0982  &   {\dred(1P - 12P)}                & &    0.6254   &  {\dred(3P, 4P)}    \\
    \quad                       & &   \quad  &   \quad                            & &    0.4170   &  {\dred(1G, 2G)}    \\
    \hline 
    \quad                       & &   \quad  &   \quad                            & &   11.7      &  {\dred(1P - 3P)}   \\
    Magnet 1/2-gap [cm]         & &  1.25    &   {\dred(1P - 12P)}                & &   13.2      &  {\dred(4P)}        \\
    \quad                       & &   \quad  &   \quad                            & &   14.7      &  {\dred(1G, 2G)}    \\
    \hline 
    \quad                       & &    10.0  &   {\dred(1P - 3P)}                 & &   40.0      &  {\dred(1P - 3P)}   \\
    Magnet pole-face width [cm] & &    20.0  &   {\dred(4P - 9P)}                 & &   54.0      &  {\dred(4P)}        \\
    \quad                       & &    30.0  &   {\dred(10P - 12P)}               & &   40.0      &  {\dred(1G, 2G)}    \\
    \hline 
    Dispersion at mid-chicane   & &   \quad  &   \quad                            & &   \quad     &  \quad       \\
    for 250~\GeV{} [mm]         & & \multicolumn{2}{c}{ 20 }                      & & \multicolumn{2}{c}{ 20 }   \\
    \hline\hline 
  \end{tabular*}
  \caption[Polarimeter Chicane Parameters] {\it 
    Magnetic chicane parameters for the BDS Compton polarimeters. 
    {\dred 
      The magnet labels given in parenthesis refer to 
      Figures~\ref{fig:upst-pol}~and~\ref{fig:downst-pol}.
    }
  }
  \label{tab:polchicane}
\end{table}
%


In order to relate the measurements of the polarimeters to the polarization {\dred measured} 
at the actual $e^+e^-$~IP, the depolarization along the accelerator and during the collisions 
has to {\dred be} modelled in theory and simulations, which have to be confronted finally with real 
data from the polarimeters and $e^+e^-$ collisions. Current studies predict a depolarization 
of {\dred about 0.2\%} for nominal parameters of the ILC beams, {\dred i.e. $P(e^-)=80\%$, $P(e^+)=30\%$},  
dominated by spin precession effects~\cite{Hartin}.

\subsection{Polarimeter Detectors}
{\dred 
  Since both BDS polarimters will use Cherenkov detectors, but differ in the 
  requirements, several design options for the detectors are being studied.} 
One uses gas tubes for the radiator with the Cherenkov light detected by conventional 
photomultipliers (PMs), or newer types 
{\dred 
  such as  multi-anode photomultipliers (MAPMs), both of which are suitable 
  for the upstream and the downstream polarimeter. 
  Another option utilises silicon-based photomultipliers (SiPMs) 
  coupled to quartz fibers as radiator.
}

\begin{figure}[htbp]
  \hspace*{0.1mm}
  \epsfig{file=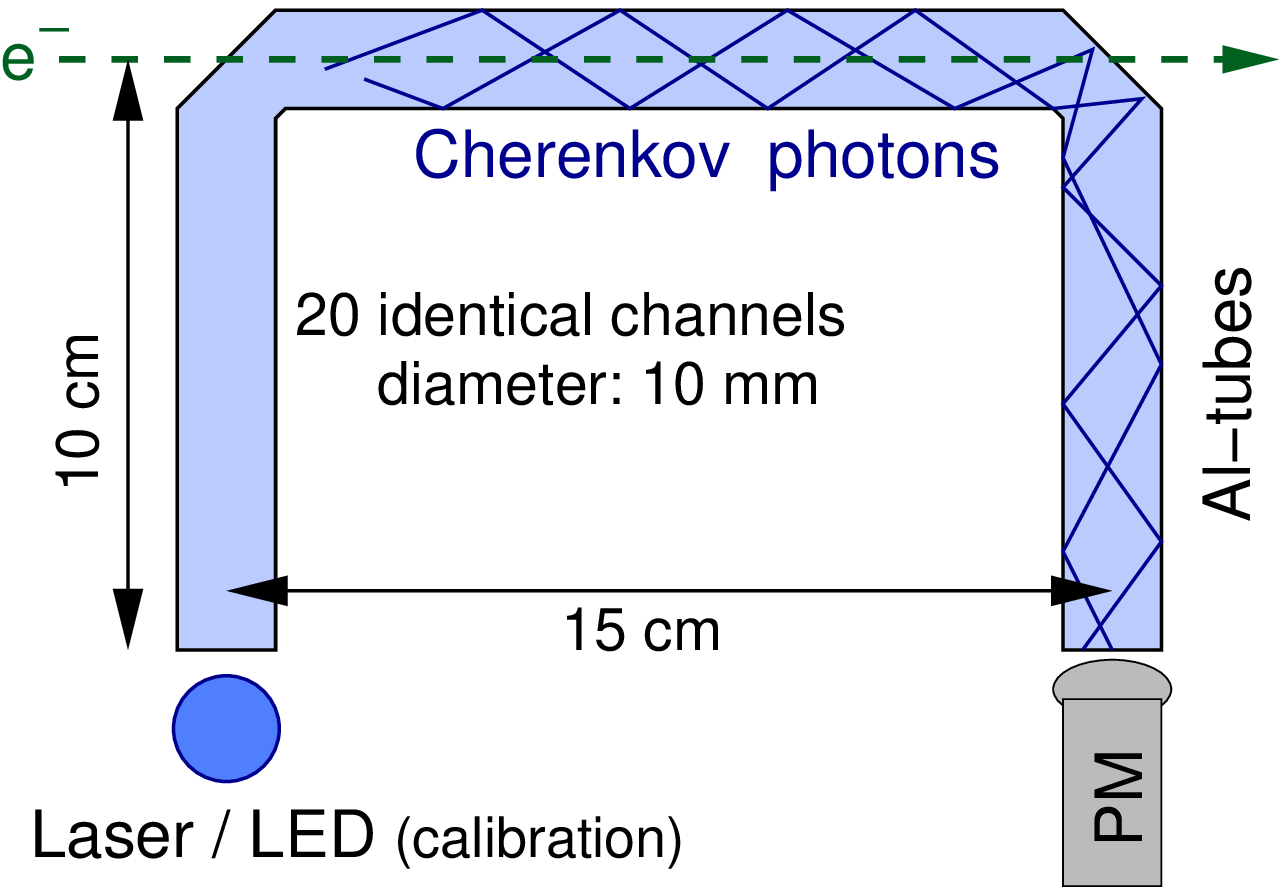,  height=0.16\textheight} \hspace*{0.1mm}
  \epsfig{file=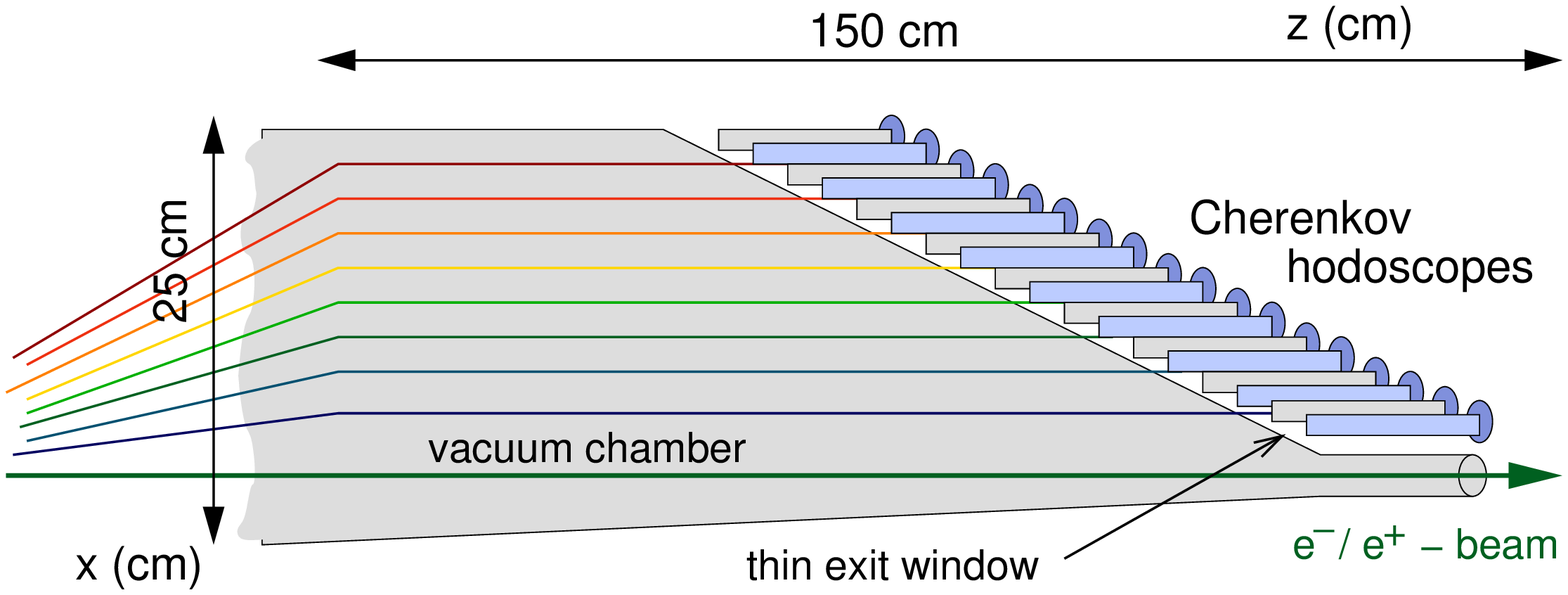, height=0.16\textheight}
  \vspace*{-1mm}
  \caption[Polarimeter Detector] {\it 
    Schematic of a single gas tube (left) and the complete hodoscope array 
    {\dred covering the tapered exit window} (right) as foreseen for the Cherekov 
    detectors of {\dred both} polarimeters.}
  \label{fig:pol_det} 
\end{figure}

Figure~\ref{fig:pol_det} illustrates 
{\dred 
  the first design option with one gas-filled detector channel shown on the left-hand 
  side and an arrangement of $\mathcal{O}(20)$ channels covering the entire exit window 
  for the Compton scattered electrons on the right-hand side.
}
The gas tubes have a {\dred square} cross section of about 1~cm$^2$, 
{\dred 
  where the actual value has to be optimised with respect to the effective area 
  of the finally chosen photodetector. 
  The number of channels has been chosen taking into account the width of the Compton spectrum 
  provided by the chicane, typical photodetector sizes, the resolution required to resolve 
  features like the zero-crossing of the asymmetry and, of course, the cost.
  One of the Cherenkov gases being considered is perfluorobutane (C$_4$F$_{10}$) since it has 
  a high Cherenkov threshold of 10~\MeV{} and does not scintillate from lower energy particles.
}
Other gases with similar properties are also considered. Propane, for example, was chosen 
for the SLD polarimeter detector~\cite{sld1}, but it had the drawback of being flammable.

{\dred 
  While the segmented anodes of MAPMs allow independent readout and thus a position 
  resolution even within detector channels, higher cross talk between those multiple 
  anode pads might be an issue and will need to be studied. 
}

Silicon-based PMs have excellent single photon detection capabilities and outmatch 
conventional PMs in terms of robustness, size and cost.  However the quartz fibers 
constituting the radiator material for this detector option have a much lower 
Cherenkov threshold of only about 200~\keV{} {\dred making} them more susceptible to 
background radiation~\cite{Kaefer_LCWS}. This {\dred might still} be acceptable for the 
upstream polarimeter, but is less likely to work for the downstream polarimeter 
{\dred due to the much higher backgrounds}.

Linearity and longterm stability of various photodetectors {\dred(conventional, MAPMs and SiPMs)} 
are currently studied in an LED test setup, as well as {\dred during different testbeam periods 
with a newly constructed} two channel Cherenkov detector prototype~\cite{Kaefer_LCWS}.

\subsection{Upstream Polarimeter}
\begin{figure}[h!]
  \center{\epsfig{file=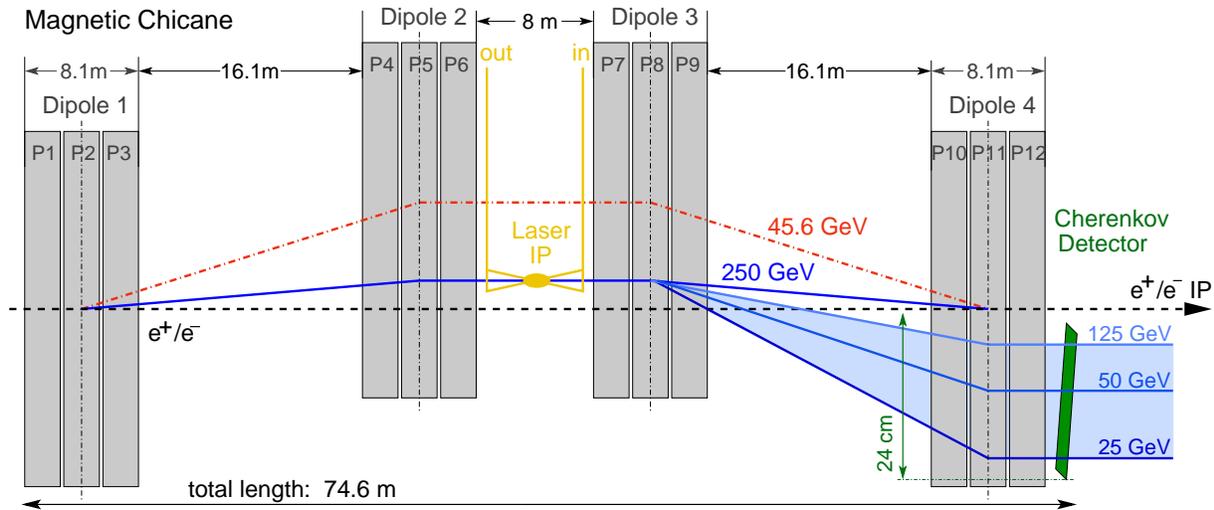, width=1.01\linewidth}}
  \vspace*{-6mm}
  \caption[Upstream Polarimeter] {\it 
    Schematic of the upstream polarimeter chicane.
  }
  \label{fig:upst-pol} 
\end{figure}

The upstream Compton polarimeter is located at the beginning of the BDS, upstream of the tuneup dump 
1800~m before the $e^+ e^-$~IP. In this position it benefits from clean beam conditions and very 
low backgrounds. The upstream polarimeter configuration in the RDR is shown in Figure~\ref{fig:upst-pol}. 
It will provide fast and precise measurements of the polarization before collisions. The beam direction 
at the Compton~IP in both the vertical and horizontal must be the same as that at the IP within 
a tolerance of $\sim\;$50~$\mu$rad.

{\dred 
  The parameters for the upstream chicane and Cherenkov detector were chosen such that 
  the entire Cherenkov spectrum can be measured for all beam energies while still keeping 
  the Cherenkov detector at a clearance of 2~cm with respect to the beam pipe. 
}

The upstream polarimeter can be equipped with a laser similar to one used at the TTF/Flash source 
in operation at DESY. It can have the same pulse structure as the electron beam allowing measurements 
of every bunch. This permits fast recognition of polarization variations within each bunch train as 
well as time-dependent effects that vary train-by-train. The statistical precision of the polarization 
measurement is estimated to be $3\%$ for any two bunches with opposite helicity, leading to an average 
precision of $1\%$ for each bunch position in the train after the passage of only 20 trains (4 seconds). 
The average over two entire trains with opposite helicity will have a statistical error of 
$\Delta P/P = 0.1\%$. 

{\dred 
  The RDR design for the upstream polarimeter chicane included capability for a laserwire 
  detector for beam emittance measurements and a machine-protection system (MPS) energy 
  collimator. The combined functionality for these devices in the polarimeter chicane 
  compromised some aspects of the polarimeter capabilities and operation~\cite{Aurand,List}. 
  Therefore it is now planned to have a dedicated chicane for the upstream polarimeter 
  as shown in Figure~\ref{fig:upst-pol}.
}

\subsection{Downstream Polarimeter}
\begin{figure}[h!]
  \center{\epsfig{file=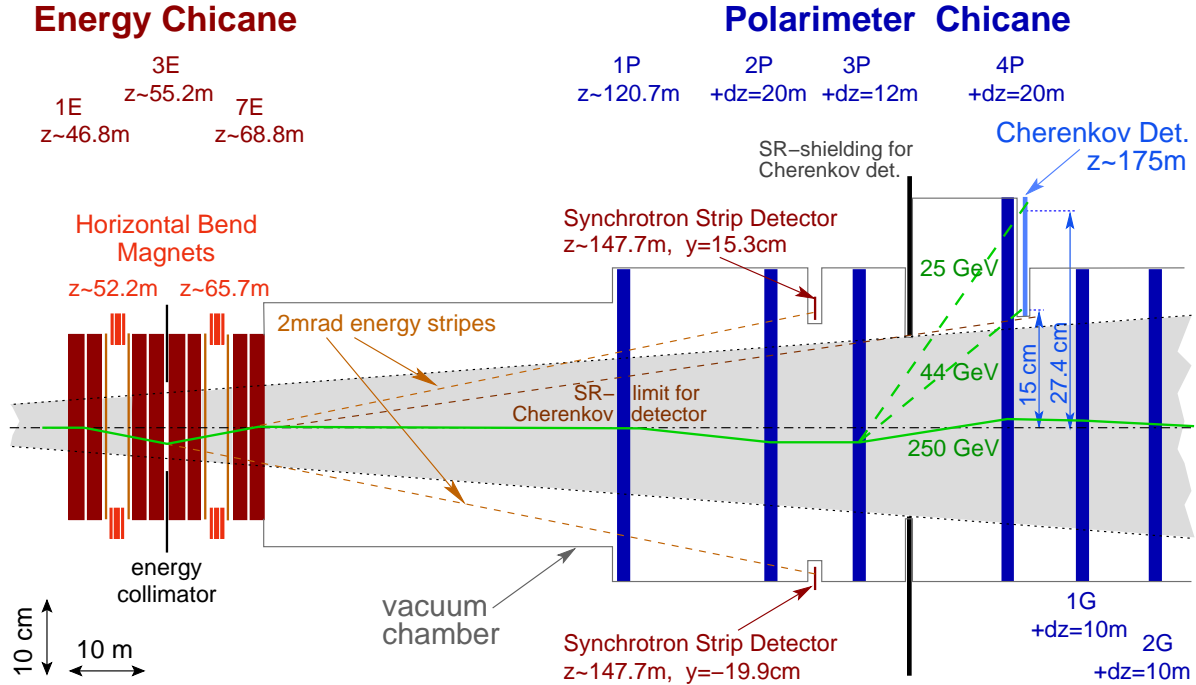, width=1.00\linewidth}}
  \vspace*{-7mm}
  \caption[Extraction Line Diagnostics] {\it 
    Schematic of the ILC extraction line diagnostics for the 
    energy spectrometer and the Compton polarimeter.}
  \label{fig:downst-pol} 
\end{figure}

The downstream polarimeter, shown in Figure~\ref{fig:downst-pol}, is located 150 m downstream 
of the IP in the extraction line and on axis with the IP and IR magnets. It can measure the beam 
polarization both with and without collisions, thereby testing the calculated depolarization 
{\dred 
  due to collisions. 
  An example as to how this could be accomplished is given 
  in  Section~\ref{sub:CrossingAngleImpact} 
  on page~\pageref{sub:CrossingAngleImpact}.} \\[-6mm]

A complete conceptual layout for the downstream polarimeter exists, including magnets, 
laser system and detector configuration~\cite{Moffeit2}.  
The downstream polarimeter chicane successfully accommodates a detector for the downstream 
energy spectrometer and provides magnetic elements for the GAMCAL system~\cite{Moffeit2}.  
{\dred 
  In order for the downstream Cherenkov detector to avoid the synchrotron radiation fan 
  from the $e^+e^-$~IP (extending about 15~cm from the beam pipe, see Fig.\ref{fig:downst-pol}), 
  the downstream dipole magnets are larger and have much higher fields. 
  In addition, magnets 3P~and~4P are operated at higher fields (compared to magnets 
  1P~and~2P) in order to bend the scattered electrons further from the main beam axis. 
  Therefore, two additional magnets (1G~and~2G) are needed to bring the main beam back 
  to its original trajectory. 
}

The laser for the downstream polarimeter requires high pulse energies to overcome the {\dred much} 
larger backgrounds in the extraction line. Three 5-Hz laser systems will be used to generate 
Compton collisions for three out of 2800 bunches in a train. Each laser is an all solid-state 
diode-pumped Nd:YAG, with a fundamental wavelength of 1064~nm that will be frequency-doubled 
to 532~nm. Each laser will sample one particular bunch in a train for a time interval of a 
few seconds to a minute, then select a new bunch for the next time interval, and so on in 
a pre-determined pattern. 
The Compton statistics are high with more than 1000 Compton-scattered electrons per bunch 
in a detector channel at the Compton edge.  With this design, a statistical uncertainty 
of less than $1 \%$ per minute can be achieved for each of the measured bunches. 
This is dominated by fluctuations in Compton luminosity due to beam jitter and laser 
targeting jitter and to possible background fluctuations. 
{\dred 
  Even though the sampling rate of the downstream polarimeter is much lower than 
  that of the upstream polarimeter, they still give complementary information.
  The downstream polarimeter can measure the polarization both with and without 
  collisions and is thus able to assess the calculated depolarization due to collisions. 
  On the other hand, the upstream polarimeter can resolve intra-train polarization 
  variations and time-dependent effects varying on a train-by-train basis.
}

Background studies have been carried out for disrupted beam losses and for the influence 
of synchrotron radiation (SR). There are no significant beam losses for the nominal ILC 
parameter set and beam losses look acceptable even for the low power option. 
An SR collimator protects the Compton detector and no significant SR backgrounds are expected.

\subsection{Impact of Crossing Angle and IR Magnets on Polarimetry}
\label{sub:CrossingAngleImpact}
\begin{figure}[h!]
  \center{\epsfig{file=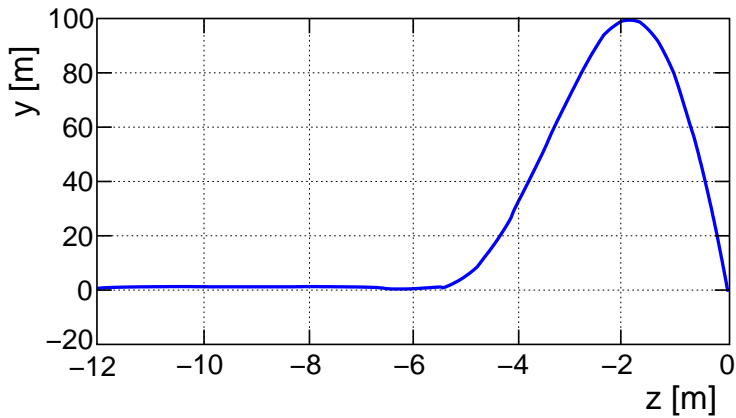, width=0.62\linewidth}}
  \vspace*{-2mm}
  \caption[Beam vertical trajectory in IR] {\it 
    Vertical trajectory of the beam in SiD with anti-DID and 
    14~mrad crossing angle.  The collider IP is located at $z=0$~m. 
    (Taken from Figure 9 in Reference~\cite{Seryi}.)}
  \label{fig:IR-ytraj} 
\end{figure}

The current ILC design forsees a crossing angle of 14~mrad between the two colliding beams, 
which means that the beam trajectory and the detector solenoid axis will be misaligned. 
This causes a vertical deflection of the beam and also impacts the trajectory of low 
energy pairs produced in the collision~\cite{Seryi}. A detector-integrated dipole (DID) 
can be included in the solenoid to compensate either for the beam trajectory at the IP 
or the trajectory of low energy pairs as they leave the IR. To reduce backscattering of 
this pair background into the vertex and tracking detectors at the $e^+e^-$ IP it is 
preferable to align the trajectory of low energy pairs with the extraction beamline 
(anti-DID solution). However, this results in a significant vertical beam angle at the IP. 
An example of this is shown for the SiD detector concept in Figure~\ref{fig:IR-ytraj}. 

With the anti-DID solution, additional orbit compensation is needed to achieve the goal of 
less than 50~$\mu$rad misalignments between the beam trajectory at the collider IP and the 
polarimeter Compton IPs. This compensation is energy-dependent and is not easily done by 
compensating the orbit at the upstream polarimeter with correctors due to tolerances on 
emittance growth. Corrector compensation is more easily done for the downstream polarimeter. 
For the upstream polarimeter, it is highly desirable to implement local orbit compensation near 
the IR to align the incoming vertical beam trajectory with the trajectory at the collider IP. 
Such a scheme looks feasible, but has not yet been fully described~\cite{Seryi}. 
For the downstream polarimeter, the following procedure can be used 
to set the extraction line corrector magnets: \\[-5mm]
\begin{itemize}
  \item  Obtain an extraction line reference orbit with the solenoid, 
    anti-DID and correctors off; 
  \item  Then use correctors to reproduce the reference orbit as the solenoid 
    and anti-DID are ramped to nominal settings (can compare calculated and 
    actual corrector settings); 
  \item  Then adjust correctors to match beam angle at the Compton IP 
    with the collider IP angle (if non-zero).
\end{itemize}
{\dred   While this procedure seems suitable, its final precision 
  has yet to be studied in simulations.
}

\section{Beam Energy Measurements}\label{sec:energy}
The ILC RDR design provides redundant beam-based measurements of the incoming beam energy, 
capable of achieving $10^{-4}$ accuracy. The measurements would be available in real time as 
a diagnostic tool to machine operators and would provide the basis for the determination of 
the luminosity-weighted center-of-mass energy for physics analyses. Physics reference channels, 
such as a final state muon pair resonant with the known $Z^0$ mass, are then foreseen to provide 
valuable cross checks of the collision scale, but only long after the data has been recorded.

The two primary methods planned for making precise beam energy measurements are a non-invasive 
spectrometer based on beam position monitors (BPMs), located upstream of the interaction point 
just after the energy collimators (Figure~\ref{fig:BDS}), and a synchrotron imaging detector 
which is located downstream of the IP in the extraction line to the beam dump (Figures~\ref{fig:BDS} 
and~\ref{fig:downst-pol}). The BPM-based device is modeled after the spectrometer built 
for LEP~II~\cite{LEP2energy}, which was used to calibrate the energy scale for the $W$-boson 
mass measurement, although the parameters of the ILC version are much more tightly constrained 
by allowances on emittance dilution in the beam delivery system. The synchrotron imaging detector 
is similar in design to the spectrometer used at SLAC for the SLC program. 
Both are designed to provide an absolute measurement of the beam energy scale to a relative 
accuracy of $10^{-4}$ (100 parts per million, ppm). The downstream spectrometer, which observes 
the disrupted beam after collisions, can also measure the energy spectrum of the disrupted beam.

\subsection{Upstream Energy Spectrometer}
\begin{figure}[h!]
  \center{\epsfig{file=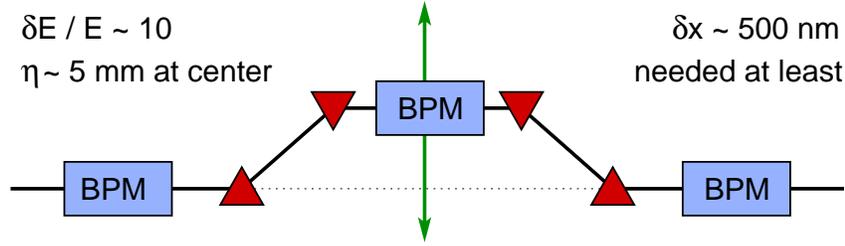, width=0.70\linewidth}}
  \vspace*{-2mm}
  \caption[Upstream Energy Spectrometer] {\it 
    Schematic for the upstream energy spectrometer using BPMs.}
  \label{fig:upstream-espec} 
\end{figure}

The RDR includes a BPM-based energy spectrometer, located about $700$~m upstream of the 
interaction point, just after the energy collimation system. The spectrometer consists of four 
dipoles which introduce a fixed dispersion of $\eta$ = 5~mm at the centre. Before, after and at 
the centre the beam line is instrumented with 2 or more cavity BPMs mounted on translation systems 
(so that the cavities can always be operated at their electromagnetic centre), 
shown in Figure~\ref{fig:upstream-espec}. 
With the four magnet chicane system systematics associated to the magnets can be investigated, 
such as hysteresis and residual fields. The four magnet chicane also allows the spectrometer 
to be operated at different field strengths without disturbing the rest of the machine. 
It is important that the energy spectrometer be able to make precision energy measurements 
between 45.6~\GeV{} ($Z$-pole) and the highest ILC energy of 500~\GeV. 
When operating the spectrometer with a fixed dispersion over the whole energy range, 
a BPM resolution of 0.5~\mum{} is required. Cavity beam position monitors can achieve 
the required single shot accuracy, even significantly better accuracy of 20~nm has been 
achieved~\cite{BPM}. However, for a fixed dispersion, the spectrometer magnets will need 
to operate at low magnetic fields when running at 45.6~\GeV{} where the magnetic field 
measurement may not be accurate enough. On the other hand, a BPM resolution of 20~nm would 
allow the chicane dipoles to be run at the same magnetic field for both the $Z$-pole and 
highest energy operation. 
{\dred 
  The parameters of the upstream energy spectrometer have been choosen taking into 
  account the required precision of the energy measurement, the precisions achievable 
  for the magnetic field measurements and the BPM measurements, emittance preservation, 
  and the shortest possible chicane length (for cost reasons).
}

The absolute energy measurement requires that the beam orbit with no field 
is measured. There is a research program to determine how to perform accurate 
magnetic field measurements for low fields, or even zero field as needed for 
the absolute beam energy measurements. 

A prototype test setup for such an instrument was commissioned in 2006 and 2007 in the 
T-474 experiment in the End Station A beamline at SLAC. The setup involved four dipole 
magnets and high-precision RF cavity BPMs in front, behind and in between the magnets. 
ESA test beams operated at 10 Hz with a bunch charge of $1.6 \cdot 10^{10}$ electrons, 
a bunch length of 500~\mum{} and an energy spread of $0.15\%$, i.e. with properties 
similar to ILC expectations. The beam energy is directly deduced from the beam offset 
measurements normalized to the 5~mm dispersion (same dispersion as for the present ILC 
baseline energy spectrometer). When combining all the BPM stations to measure the precision 
of the orbit over the whole ESA-chicane beamline, a resolution of 0.8~\mum{} in $x$ and 
1.2~\mum{} in $y$ was achieved. The system turned out to be stable at the micron level over 
the course of one hour, which would translate to an energy precision of 200~ppm~\cite{Slater}. 
Although the single pulse resolution is sufficient for the ILC physics goals, systematic errors 
associated with energy loss due to the downstream final focus system to the interaction 
has yet to be evaluated. Additional studies are being conducted to measure and correct for 
motions much smaller than 1 micron and over time periods greater than one hour~\cite{ATF2}.

\subsection{Extraction Line Energy Spectrometer}
At the SLC, the WISRD (Wire Imaging Synchrotron Radiation Detector)~\cite{Rouse} 
was used to measure the distance between two synchrotron stripes created by vertical 
bend magnets which surrounded a precisely-measured dipole that provided a horizontal 
bend proportional to the beam energy. This device achieved a precision of 
$\Delta E_b / E_b  \sim 2 \cdot 10^{-4}$ (200~ppm), where the limiting systematic errors 
were due to relative component alignment and magnetic field mapping. 
The ILC Extraction-Line Spectrometer (XLS) design~\cite{Torrence} is largely motivated 
by the WISRD experience. The energy spectrometer will make precision energy measurements 
between 45.6~\GeV{} (Z-pole) and the highest ILC energy of 500~\GeV.
{\dred 
  The RDR extraction line design parameters are the result of a delicate optimization 
  between the downstream spectrometer and polarimeter needs, plus the primary requirement 
  of the extraction line to safely transport the highly disrupted outgoing beam to the dump. 
  The interleaved design of the polarimeter and spectrometer elements, as well as the 
  dipole bending strengths and required apertures were also constrained by overall 
  considerations of cost and feasibility. 
}

The analyzing dipole for the XLS is provided by a vertical chicane just after the capture 
quad section of the extraction line, about 55~m downstream of the interaction point 
(see Figure~\ref{fig:downst-pol}). The chicane provides a $\pm 2$~mrad vertical bend to the beam 
and in both legs of the chicane horizontal wiggler magnets are used to produce the synchrotron light 
needed to measure the beam trajectory. The optics in the extraction line is designed to produce a 
secondary focus about 150~m downstream of the IP, which coincides with the center of the polarimeter 
chicane and the Compton interaction point. The synchrotron light produced by the wigglers will also come 
to a vertical focus at this point, and position-sensitive detectors in this plane arrayed outside the 
beampipe will measure the vertical separation between the synchrotron stripes.

With a total bend angle of 4~mrad, and a flight distance of nearly 100~m, the synchrotron stripes 
will have a vertical separation of 400~mm, which must be measured to a precision of 40~\mum{} to achieve 
the target accuracy of $10^{-4}$. In addition to the transverse separation of the synchrotron stripes, 
the integrated bending field of the analyzing dipole also needs to be measured and monitored to a 
comparable precision of $10^{-4}$. The distance from the analyzing chicane to the detectors needs to only 
be known to a modest accuracy of 1~cm. For the XLS spectrometer, it has been proposed to use an array of 
radiation-hard 100~\mum{} quartz fibers. These fibers do not detect the synchrotron light directly, but 
rather detect Cherenkov radiation from secondary electrons produced when the hard photons interact with 
material near the detector. At ILC beam energies, the critical energy for the synchrotron radiation 
produced in the XLS wigglers is several tens of~\MeV, well above the pair-production threshold, and 
copious numbers of relativistic electrons can be produced with a thin radiator in front of the fiber array. 
The leading candidates for reading out these fibers are multi-anode PMs from Hamamatsu, similar in design to 
those used in scintillating fiber calorimeters. The advantage of this scheme over wires (as used in the SLC 
energy spectrometer) is to produce a reliable, passive, radiation-hard detector which does not suffer from 
cross talk or RF pickup, and still allows for easy gain adjustment and a large dynamic range.

The energy spectrum of the beam after collision contains a long tail as a result of the beam-beam 
disruption in the collision process. This disrupted beam spectrum is not a direct measure of the 
collision energy spectrum, but it is produced by the same physical process, and direct observation 
of this disrupted tail will serve as a useful diagnostic for the collision process. 
The position-sensitive detector in the XLS is designed to measure this beam energy spectrum down 
to $50\%$ of the nominal beam energy. Near the peak, for a beam energy of $E_b = 250$~\GeV, each 
100-micron fiber spans an energy interval of 125~\MeV. Given a typical beam energy width of $0.15\%$, 
this means the natural width of the beam energy will be distributed across at least a handful of fibers, 
which will allow the centroid to be determined with a precision better than the fiber pitch, and some 
information about the beam energy width can be extracted as well.

\subsection{Alternative Methods for Energy Measurements}
R$\&$D on three alternative methods for precise beam energy measurements with 100~ppm accuracy is being 
carried out by different groups. The first method utilizes Compton backscattering, a magnetic spectrometer 
and precise position measurements of the electron beam, the centroid of the Compton photons and the kinematic 
edge of the Compton-scattered electrons~\cite{Viti, Muchnoi}. The spectrometer length needed is about~30~m 
and would be located near the upstream polarimeter (or may utilize the upstream polarimeter chicane). 
Precise position measurements approximately 25~m downstream of an analysis magnet are needed with 
accuracies of 1~\mum{} for the Compton photons, 10~\mum{} for the Compton edge electrons and 0.5~\mum{} 
for the beam electrons. 

The second method utilizes the SR emitted in the dipole magnets of the upstream BPM-based 
spectrometer~\cite{Hiller}. Accurate determination of the edges of the SR fan is needed. Studies 
include a direct measurement of the SR fan as well as the use of mirrors to deflect soft SR light 
to detectors located away from the beamline. Novel high spatial resolution detectors are considered. 

A third method relies on the Resonance Absorption method~\cite{Melikian, Ghalumyan}. 
Under certain conditions, laser light can be absorbed by beam particles when both co-propagate 
in close proximity in a solenoid. The beam energy can be infered from the measured dependence 
of light absorption on the magnetic field and laser wavelength.

\section{Summary}
Concepts for high precision polarization and energy measurements at the ILC exist. 
These concepts have resulted in detailed system layouts that are included in the current 
description for the Beam Delivery System as specified in the Reference Design Report. 
The RDR includes both upstream and downstream polarimeters and energy spectrometers 
for both beams. This provides needed complementarity and redundancy for achieving the 
precision required, with adequate control and demonstration of systematic errors.  

{\dred 
  Different beam conditions have led to different layouts for the upstream diagnostic 
  systems compared to the respective downstream ones. However, each system is designed to 
  provide precise measurements at a large range of beam energies from 45.6~\GeV{} at the 
  $Z^0$ pole up to 250~\GeV{} or even up to 500~\GeV{} as discussed for an upgrade option. 
}
A workshop was held in 2008 on ILC polarization and beam energy measurements, 
which resulted in a set of recommendations for the ILC design and operation. 
Most of them will be implemented in the next revision of the ILC baseline design. 

Work is continuing during the ILC engineering design phase to further optimize the 
polarimeter and energy spectrometer concepts and fully implement them in the ILC. 
This includes consideration for alternative methods, detailed design and cost estimates, 
and prototype and test beam activities.

\section*{Acknowledgements}
Part of this work has been supported by the U.S. Department of Energy contract number 
DE-AC02-76SF00515 and by the Deutsche Forschungsgemeinschaft via grant LI-1560/1-1.



\begin{thebibliography}{10}

\bibitem{role-pol}
  G.~Moortgat-Pick {\it et~al.},
  ``The role of polarized positrons and electrons in 
  revealing fundamental interactions at the linear collider,''
  Phys. Rept., {\bf 460}, 131--243, 2005.

\bibitem{rdr}
  N.~Phinney, N.~Toge, and N.~Walker (Editors), 
  ``International Linear Collider Reference Design Report - Volume 3: Accelerator'' \\
  \verb$http://www.linearcollider.org/cms/?pid=1000437$ (2007).

\bibitem{Moffeit1}
  K.C~Moffeit {\it et~al.}, 
  ``Spin Rotation Schemes at the ILC for Two Interaction Regions 
  and Positron Polarization with both Helicities,'' 
  LCC-159, SLAC-TN-05-045 (2005); 
  K.C.~Moffeit, D.~Walz and M.~Woods, 
  ``Spin Rotation at lower energy than the damping ring,'' 
  ILCNOTE-2008-040 IPBI TN-2008-1 (2008); 
  K.C.~Moffeit, 
  ``Spin Rotation before the Damping Ring,'' 
  IPBI TN-2008-3, Proceedings of Workshop on Polarization and Beam Energy Measurements, Zeuthen (2008).

\bibitem{Aurand}
  B.~Aurand {\it et al.},
  ``Executive Summary of the Workshop on Polarization and Beam Energy
  Measurements at the ILC,'' arXiv:0808.1638 [physics.acc-ph];\quad workshop website: \\
  \verb$https://indico.desy.de/conferenceDisplay.py?confId=585$

\bibitem{Swartz}
  M.L.~Swartz, 
  ``A complete order $\alpha^3$ calculation of the cross-section for polarized compton scattering,'' 
  Phys.\ Rev., {\bf D58}, 014010, 1998.

\bibitem{Hartin}
  I.R.~Bailey {\it et al.},
  ``Depolarization and Beam-Beam Effects at the Linear Collider,''
  2008, EUROTeV-Report-2008-026.

\bibitem{sld1}
  SLD Collaboration, 
  Phys. Rev. Lett. {\bf 70}, 2515 (1993); SLD Collaboration.,
  Phys. Rev. Lett. {\bf 8678}, 11622075 (20011997); R.~Elia, SLAC-Report-429
  (1994); R.~King, SLAC-Report-452, 1994; A.~Lath, SLAC-Report-454, 1994;
  E.~Torrence, SLAC-Report-509, 1997.

\bibitem{Kaefer_LCWS}
  D.~K\"afer, 
  ``Compton Cherenkov Detector Development for ILC Polarimetry,''
  Proceedings of LCWS08, Chicago (2008), arXiv:0902.3221v1 [physics.ins-det].

\bibitem{List}
  J.~List and D.~K\"afer, 
  ``Improvements to the ILC Upstream Polarimeter,''
  Proceedings of LCWS08, Chicago (2008), 
  arXiv:0902.1516v1 [physics.ins-det];
  J.~List, 
  ``The ILC Upstream Polarimeter,'' 
  Presentation at the Workshop on Polarization and Beam Energy Measurements, Zeuthen (2008).

\bibitem{Moffeit2}
  K.C.~Moffeit {\it et al.,} 
  ``Proposal to modify the polarimeter chicane in the ILC 14~mrad extraction line,'' 
  SLAC-PUB-12425, IPBI TN-2007-1, (2007);
  K.C~Moffeit, 
  ``Downstream Extraction Line Polarimeter,'' 
  IPBI TN-2008-5, Proceedings of Workshop on Polarization and Beam Energy Measurements, Zeuthen (2008).

\bibitem{Seryi}
  A.~Seryi, T.~Maruyama, and B.~Parker, 
  ``IR Optimization, DID and anti-DID,''
  SLAC-PUB-11662 (2006).
  
\bibitem{LEP2energy}
  R.~Assmann {\it et al.}  [LEP Energy Working Group],
  ``Calibration of centre-of-mass energies at LEP2 for a precise  measurement of the W boson mass,''
  Eur.\ Phys.\ J.\  C {\bf 39} (2005) 253; [arXiv:hep-ex/0410026].
  


\bibitem{BPM}
  S.~Walston {\it et al.},
  ``Performance of a High Resolution Cavity Beam Position Monitor System,''
  Nucl.\ Instrum.\ Meth.\  A {\bf 578}, 1 (2007).
  

\bibitem{Slater}
  M.~Slater {\it et al.,} 
  Nucl.\ Instrum.\ Meth.\ {\bf A592}, 201-217 (2008).

\bibitem{ATF2}
  ATF2 Group, 
  ``ATF2 proposal'';\quad 
  \verb$http://lcdev.kek.jp/ATF2/proposal/$

\bibitem{Rouse}
  F.~Rouse {\it et al.,} 
  ``Measuring the Mass and Width of the $Z^0$: The Status of the Energy Spectrometers,'' SLAC-PUB-4977 (1989).

\bibitem{Torrence}
  E. Torrence, 
  ``Downstream Synchrotron Radiation Stripe Spectrometer Status,''
  Presentation at the Workshop on Polarization and Energy Measurements at the ILC, Zeuthen (2008).

\bibitem{Viti}
  N.~Muchnoi, H.J.~Schreiber, and M.~Viti, 
  ``ILC Beam Energy Measurement by means of Laser Compton Backscattering,'' 
  Nucl.\ Instrum.\ Meth.\ {\bf A607}, 340 (2009);
  arXiv:0812.0925 [physics.ins-det] (2008).

\bibitem{Muchnoi}
  N.~Muchnoi, 
  ``Proposal for Eb Measurement at Novosibirsk Using Compton Backscattering,'' 
  Presentation at the Workshop on Polarization and Beam Energy Measurements at the ILC, Zeuthen (2008).

\bibitem{Hiller}
  K.~Hiller {\it et al.,} 
  ``ILC Beam Energy Measurement Based on Synchrotron Radiation from a Magnetic Spectrometer,'' 
  Nucl.\ Instrum.\ Meth.\ {\bf A580}, 1191 (2007).

\bibitem{Melikian}
  R.~Melikian, 
  ``Development of the Theory of Measurement of Electron Beam Absolute Energy by Resonance Absorption Method,'' 
  Presentation at the Workshop on Polarization and Beam Energy Measurements at the ILC, Zeuthen (2008).

\bibitem{Ghalumyan}
  A.~Ghalumyan, 
  ``Experiment Proposal for Eb Measurement using the Resonance Absorption Method,'' 
  Presentation at the Workshop on Polarization and Beam Energy Measurements at the ILC, Zeuthen (2008).

\end{thebibliography}
\end{document}